# Fast View Frustum Culling of Spatial Object by Analytical Bounding Bin

Munsu Ju [a], Yunchol Jong [a]

[a]Center of Natural Science, University of Sciences, Pyongyang, DPR Korea

**Abstract.** It is a common sense to apply the VFC (view frustum culling) of spatial object to bounding cube of the object in 3D graphics. The accuracy of VFC can not be guaranteed even in cube rotated three-dimensionally. In this paper is proposed a method which is able to carry out more precise and fast VFC of any spatial object in the image domain of cube by an analytic mapping, and is demonstrated the effect of the method for terrain block on global surface.

## 1. Introduction

The test of intersection to VF(view frustum) is indispensable process in rendering of 3D object. In the traditional method using bounding box, this test takes about 1/3~1/2 portion of total processing time [2, 3]. Generally, the test is made about six sides forming VF. There have been proposed several methods: one which makes use of minimum of distance with sign from 8 corner points of bounding cube of the object to each section [2]; one which does not test for some sides in view of geometrical feature of the spatial object [4,5]; one which reduces a computational complexity by making use of a spheroid as bounding domain of spatial object [6, 7]; one which reduces the complexity by test of relation of view angle and coning angle, using a conoid from view point to object as dynamic bounding domain of spatial object; and one which mixes bounding sphere and bounding cone [8]. All these methods extend bounding domain of spatial object or take account of bounding domain of VF, and are helpful to primary fast test of objects which goes beyond sight. But, these methods are not any help to make precise testing whether the object intersects boundary sides of VF or lies completely in the inside of VF.

The problem is to make the size of bounding domain small as possible and to reduce computational complexity of test. The smallest bounding domain is represented by image domain of a cube by analytic mapping in case of some spatial objects. For example, most of objects in CAD can be represented by an analytic mapping and the smallest bounding domain of terrain block on global ellipsoid can be represented as image of a cube of 3D-Eucleadian space, which has latitude, longitude and altitude as coordinate axes, by triangle polynomial.

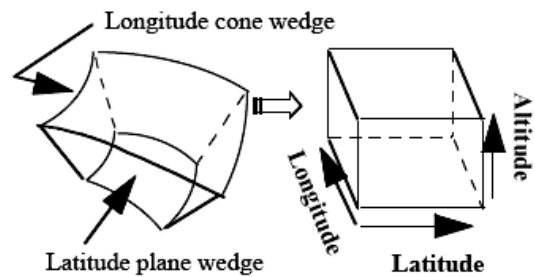

**Fig.1**. Relation between constrained bin in geocentric coordinate system and constrained bin in lat/lon/altitude Cartesian coordinate system.

We propose a fast algorithm for intersection test of this smallest bounding domain and VF. In the algorithm, the analytic property of the mapping can be weakened to twice-continuously differentiability.

## 2. Algorithm

Consider the following 3D analytic mapping $f = (f_1, f_2, f_3)^T$ :

$$f_i(x) = f_i(x_1, x_2, x_3), \quad i = 1,2,3 \qquad (1)$$

The quadratic approximation of $f$ at point $x_0$ is obtained by

$$f_i(x_0 + x) \approx f_i(x_0) + \nabla f_i(x_0)^T (x) + \frac{1}{2} x^T H_i(x_0) x, \quad i = 1,2,3, \qquad (2)$$

where $H_i(x_0)$ is Hessian matrix of $f_i$ at $x_0$. Let $g_i(x)$ denote the right-side of

$$f_i(x_0) - f_i(x_0 + x) \approx -\nabla f_i(x_0)^T (x) - \frac{1}{2} x^T H_i(x_0) x, \quad i = 1,2,3 \qquad (3)$$

and $P_j, j = 1,...,6$ be six sections of VF. Let $\bar{n}_j$ be outer normal unit vector of each $P_j$, $O$ be view point in world coordinate, and $sca_j(x)$ be scalar product of difference vector $g(x) = (g_1(x), g_2(x), g_3(x))^T$ and $\bar{n}_j$. Then $sca_j(x)$ is expressed by a quadratic polynomial of $x$. Denoting the distance taking sign into account from point $f(x_0)$ to section $P_j$ by $d_j$, the problem of intersecting with VF can be solved as follows. First, we find the maximum $\max s_j$ and the minimum $\min s_j$ of the quadratic polynomial $sca_j(x)$ for every $x$ of cube. If $\max s_j < d_j$, then the image domain of the cube completely lies in the outside of section $P_j$. And if $d_j > \min s_j$, then the image domain of the cube completely lies in the inside of section $P_j$. If $\min s_j \leq d_j \leq \max s_j$, then the image domain of the cube completely intersects the section $P_j$.

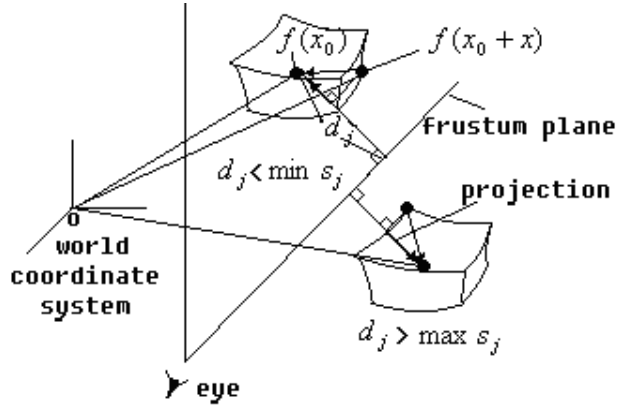

**Fig.2**. Test of intersecting of nonlinear constrained box and VF

The maximum $\max s_j$ and the minimum $\min s_j$ of the quadratic polynomial $sca_j(x)$ in the cube are attained among stationary point and eight corner points of the cube. Therefore, we take the maximum and the minimum among values of $sca_j(x)$ at the 9 points as $\max s_j$ and $\min s_j$, respectively. In practice, we extend the cube domain a little in proportion to average radius of the domain and carry out the above-mentioned test so that the influence of error due to the polynomial approximation of the mapping is weakened.

The quadratic polynomial $g_i(x)$ is a second-order approximation of the analytic function $f_i$ round $x_0$. When more precise approximation is needed, we can add higher order terms of Tailor expansion. However, solving higher-order nonlinear equations needs a lot of computational effort, and second-order approximation is appropriate for real-time implementation.

## 3. Application to real-time visualization of global terrain

The environments of our experiment are as follows: Pentium IV, 2.4GHz CPU; 512MB RAM; Windows XP OS; 1024×768 pixel$^2$ resolution. The development environments are Microsoft NET Visual Studio 5.0, QT 4.5.2 and OpenGL. The hierarchical pyramid model [5] of globe is used as global terrain and GTOP30 data [1] is used as terrain data.

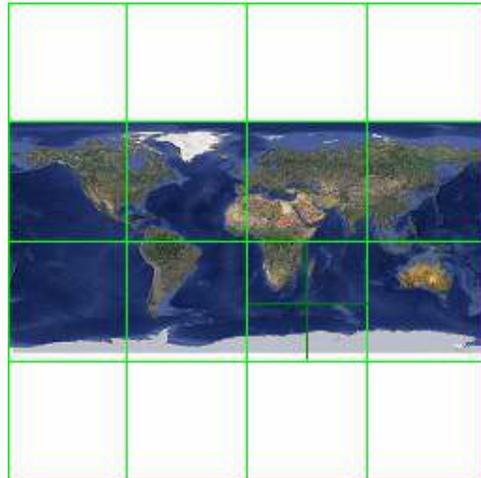

**Fig. 3**. Initial division of the earth terrain

In the fast rendering of terrain on the earth surface, recursive quadrant tile division of globe terrain block and visuality test of each tile block, i.e., test of whether the block is in the outside or in the inside of VF or intersects VF is recursively carried out. If a block is completely in the inside of sight, further visuality test for recursive quad division blocks of the block is needless. The blocks, which are completely in the outside of sight, are excluded from further consideration. Therefore, the visuality test in the recursive quad division is repeated for only blocks intersecting boundary of sight.

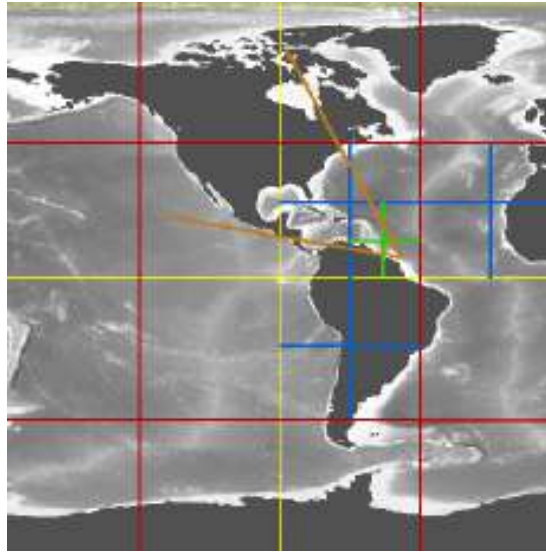
**Fig.4.** Recursive quad division of globe terrain block and visuality test

In the case that the earth surface is approximated by spheroid or ellipsoid, our algorithm gives criterion of accurate test which is free from errors in visuality test of terrain height blocks to be divided. However, the application of this algorithm to terrain height blocks requires finding in advance maximum and minimum height of terrain in each block, and cube domain in Cartesian coordinate with latitude, longitude and altitude axes is expressed by direct product of latitude interval, longitude interval and interval between minimum and maximum. The image domain of this cube by sphere triangle polynomial mapping has curved form of rectangular plate similarly to the crust cut like Fig.1.

The sphere triangle mapping representing any position in vicinity of terrain surface is as follows.

$$\begin{aligned} x &= (R+H)\cos(lat)\sin(lon) \\ y &= (R+H)\sin(lat) \\ z &= (R+H)\cos(lat)\cos(lon), \end{aligned} \quad (4)$$

where *lat, lon*, H and R is latitude, longitude, altitude of this position and average radius of the earth.

The cube in latitude-longitude-altitude coordinate system corresponding to initial division tile of Fig.3 is $[-\pi, \pi] \times [-\pi, \pi] \times [0, 9000]$.

Let $r = R+H, \varphi = lat, \theta = lon$ and $X = (r, \varphi, \theta)$, and denote the right-sides of (4) by $f_i(X), i=1,2,3$. Then, in view of (3), we have

$$sca_j(X) = g(X)^T \bar{n}^j = r1_j r + \varphi 1_j \varphi + \theta 1_j \theta + $$
$$+ \frac{1}{2}\varphi 2_j \varphi^2 + \frac{1}{2}\theta 2_j \theta^2 + A_j \varphi\theta + B_j r\varphi + C_j r\theta, \ j=1,\ldots,6,$$

where

$$g(X) = (g_1(X), g_2(X), g_3(X))^T, \ \bar{n}^j = (n_1^j, n_2^j, n_3^j), \ j=1,\ldots,6$$

$$r1_j = -\left(\cos\varphi\cos\theta\, n_1^j + \cos\varphi\sin\theta\, n_2^j + \sin\varphi\, n_3^j\right)$$
$$\varphi1_j = r\left(\sin\varphi\cos\theta\, n_1^j + \sin\varphi\sin\theta\, n_2^j - \cos\varphi\, n_3^j\right)$$
$$\theta1_j = r\left(\cos\varphi\sin\theta\, n_1^j - \cos\varphi\cos\theta\, n_2^j\right)$$
$$\varphi2_j = r\left(\cos\varphi\cos\theta\, n_1^j + \cos\varphi\sin\theta\, n_2^j + \sin\varphi\, n_3^j\right)$$
$$\theta2_j = -r\left(\cos\varphi\cos\theta\, n_1^j + \cos\varphi\sin\theta\, n_2^j\right)$$
$$A_j = -r\left(\sin\varphi\sin\theta\, n_1^j - \sin\varphi\cos\theta\, n_2^j\right)$$
$$B_j = \sin\varphi\cos\theta\, n_1^j + \sin\varphi\sin\theta\, n_2^j - \cos\varphi\, n_3^j$$
$$C_j = \cos\varphi\sin\theta\, n_1^j - \cos\varphi\cos\theta\, n_2^j$$

Then, the stationary point of $sca_j(X)$ satisfies the following equations:

$$\frac{\partial sca_j(X)}{\partial r} = r1_j + B_j\varphi + C_j\theta = 0,$$
$$\frac{\partial sca_j(X)}{\partial \varphi} = \varphi1_j + \varphi2_j\varphi + B_j r = 0,$$
$$\frac{\partial sca_j(X)}{\partial \theta} = \theta1_j + \theta2_j\theta + C_j r = 0$$

Solving these equations, we obtain the stationary point $\overline{X} = (\overline{r}, \overline{\varphi}, \overline{\theta})$, where

$$\overline{\varphi} = \frac{-C_j(C_j\varphi1_j - B_j\theta1_j) - B_j\theta2_j r1_j}{C_j^2\varphi2_j + B_j^2\theta2_j},$$
$$\overline{\theta} = \frac{B_j(C_j\varphi1_j - B_j\theta1_j) - C_j\varphi2_j r1_j}{C_j^2\varphi2_j + B_j^2\theta2_j},$$
$$\overline{r} = \frac{-(\theta1_j - \theta2_j\theta1_j)}{C_j}.$$

Our rendering algorithm carries out recursive quad division having this tile as 0-level and chooses 128 tiles obtained in the recursive divisions of 4-level as starting blocks of test. In the latitude-longitude-altitude coordinate system corresponding to these blocks, the lengths of latitude and longitude direction of the cube are $0.0625\pi$. We take cube of 1.1 times as large as original size in each step of visuality test and find quadratic polynomial approximation of sphere triangle mapping defined by (4), and carry out visuality test according to our algorithm. We compared our algorithm with the visuality test algorithm [2] of 8 corner points of terrain block in world coordinate. In the algorithm of [2], the intersection of the block and sight in both edges of side close to VF was misjudged as lying in the outside of sight. However, our algorithm overcame this phenomenon and reduced the amount of blocks by about 50 % in the recursive division processes.

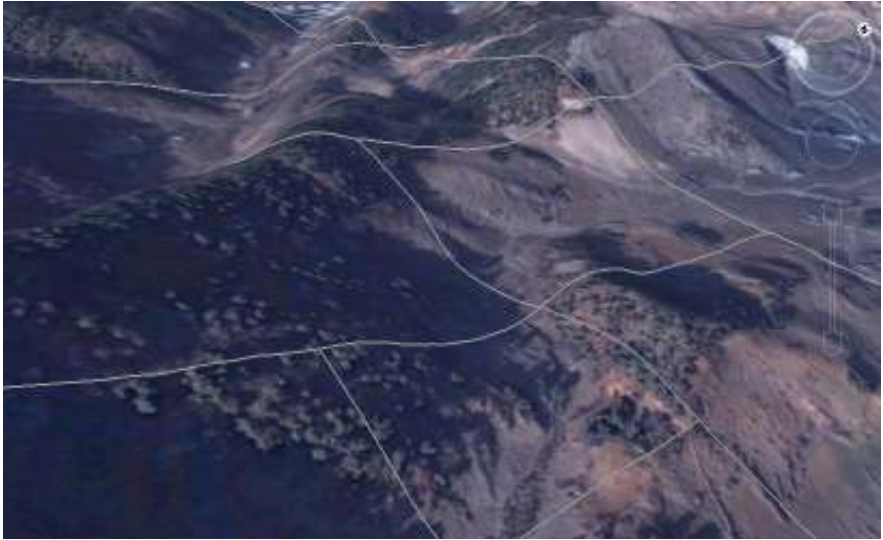

**Fig.5**. Rendering of terrain blocks

## 4. Experiment and comparison

The following figures show the performances of experiment results by the method of [2] and our method.

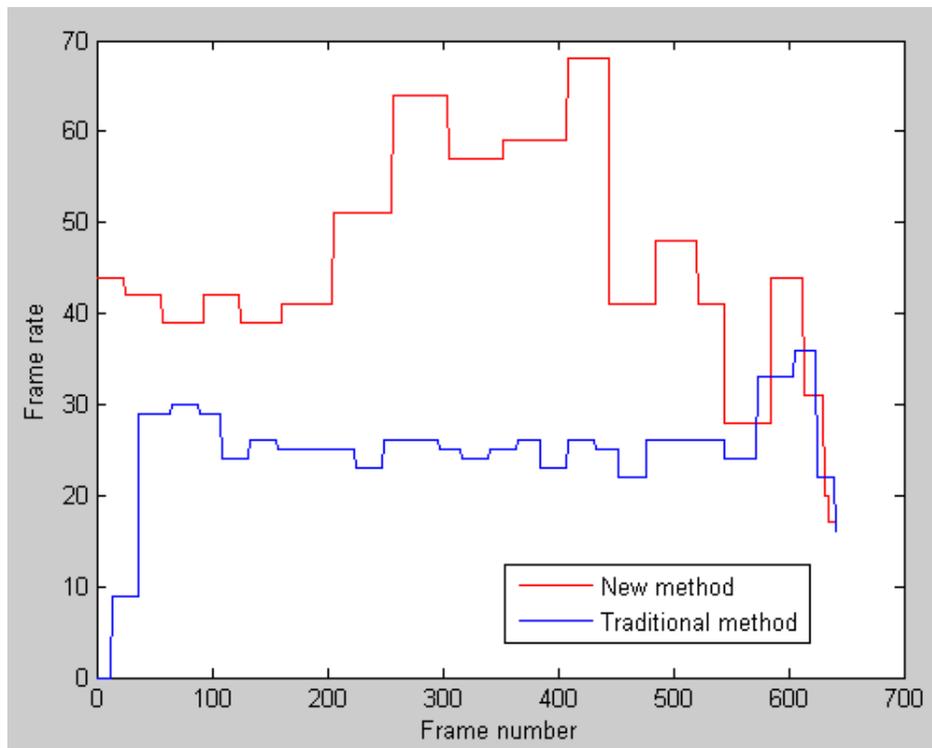

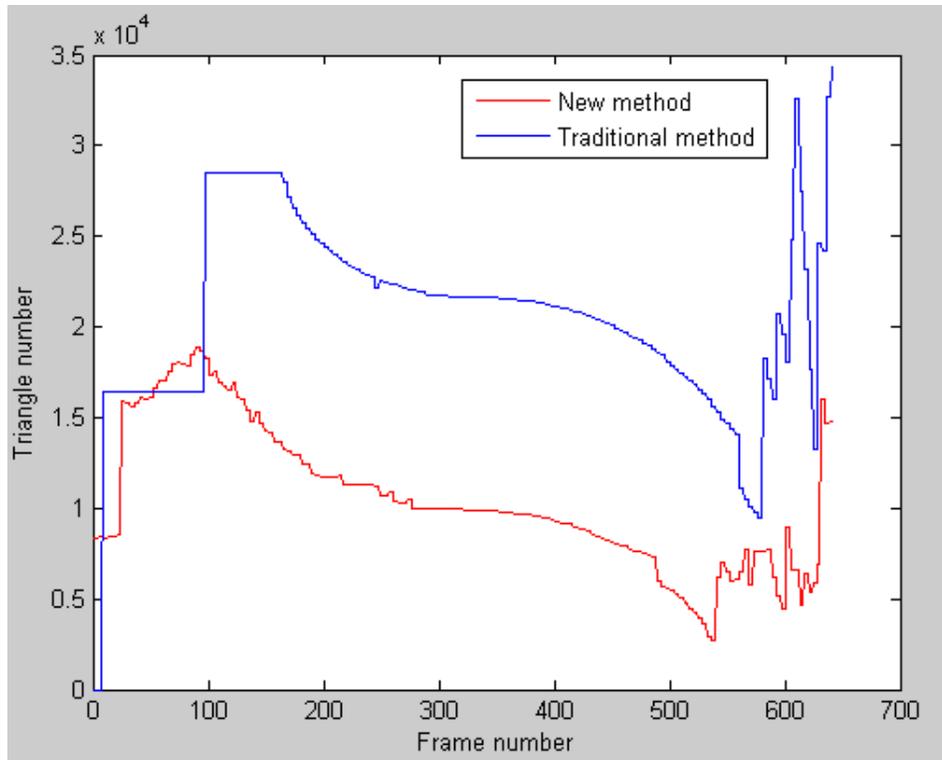

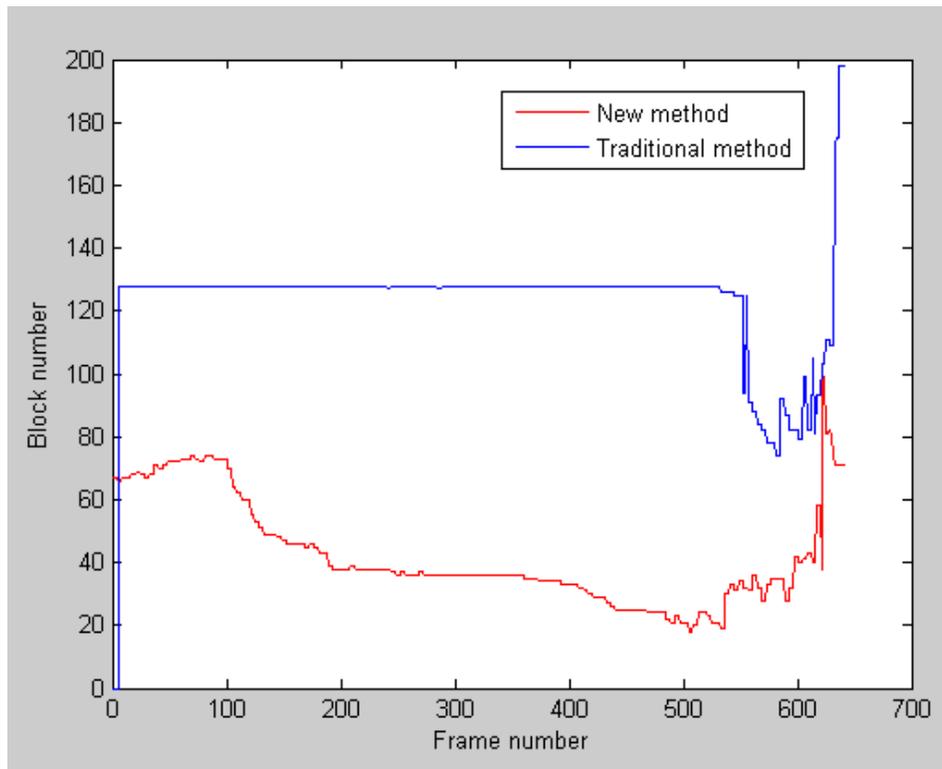

As you see from the figures, our method is much better than one of [2].

## 5. Conclusion

This paper has proposed an algorithm for elaborated and fast VFC that is primary and essential in 3D graphics, and has demonstrated the effectiveness of the algorithm in real-time visualization of global terrain. The main idea of the algorithm is to use second-order approximation of curve-transformed 3D cube (bin) by Tailor expansion. The main feature of the algorithm is to ensure both the correctness and speed of VFC. We expect that our algorithm can be applied to not only terrain rendering, but also almost all fields of 3D graphics because of its generalization ability.

---

E-mail address: yuncholjong@yahoo.com